# RARD: The Related-Article Recommendation Dataset

JOERAN BEEL, Trinity College Dublin, Department of Computer Science, ADAPT Centre, Ireland
ZELJKO CAREVIC, GESIS – Leibniz Institute for the Social Sciences, Germany
JOHANN SCHAIBLE, GESIS – Leibniz Institute for the Social Sciences, Germany
GABOR NEUSCH, Corvinus University of Budapest, Department of Information Systems, Hungary

Recommender-system datasets are used for recommender-system offline evaluations, training machine-learning algorithms, and exploring user behavior. While there are many datasets for recommender systems in the domains of movies, books, and music, there are rather few datasets from research-paper recommender systems. In this paper, we introduce RARD, the Related-Article Recommendation Dataset, from the digital library Sowiport and the recommendation-as-a-service provider Mr. DLib. The dataset contains information about 57.4 million recommendations that were displayed to the users of Sowiport. Information includes details on which recommendation approaches were used (e.g. content-based filtering, stereotype, most popular), what types of features were used in content based filtering (simple terms vs. keyphrases), where the features were extracted from (title or abstract), and the time when recommendations were delivered and clicked. In addition, the dataset contains an implicit item-item rating matrix that was created based on the recommendation click logs. RARD enables researchers to train machine learning algorithms for research-paper recommendations, perform offline evaluations, and do research on data from Mr. DLib's recommender system, without implementing a recommender system themselves. In the field of scientific recommender systems, our dataset is unique. To the best of our knowledge, there is no dataset with more (implicit) ratings available, and that many variations of recommendation algorithms. The dataset is available at http://data.mr-dlib.org, and published under the "Creative Commons Attribution 3.0 Unported (CC-BY)" license.

## 1 INTRODUCTION

Recommender-system datasets are available in many disciplines such as movies [Harper and Konstan 2016], books [Ziegler et al. 2005], and music [Bertin-Mahieux et al. 2011]. The datasets can be used for offline evaluations, to train machine learning algorithms, and to explore the dataset-provider's users and items. The interest in such datasets may be immense: one of the most popular datasets, MovieLens, was downloaded 140,000 times in 2014 [Harper and Konstan 2016], and Google Scholar lists 10,600 articles that mention the MovieLens dataset (**Figure 1**).

In this paper, we introduce a new dataset named *RARD* – the *R*elated-*A*rticle *R*ecommendation *D*ataset. RARD is based on the research-paper recommender system Mr. DLib[1] [Beel, Aizawa, et al. 2017], and the digital library Sowiport[2] [Hienert et al. 2015]. The dataset contains the delivery and click logs for 57 million displayed recommendations including information about the recommendation algorithms, as well as an implicit item-item ratings matrix based on the logs. The recommendation logs can be used to analyze how effective the recommendation algorithms and parameters are, and to write and publish research papers about the findings. The rating matrix may be used to train machine learning recommendation algorithms, and conduct offline evaluations.

---

[1] http://mr-dlib.org

[2] http://sowiport.gesis.org

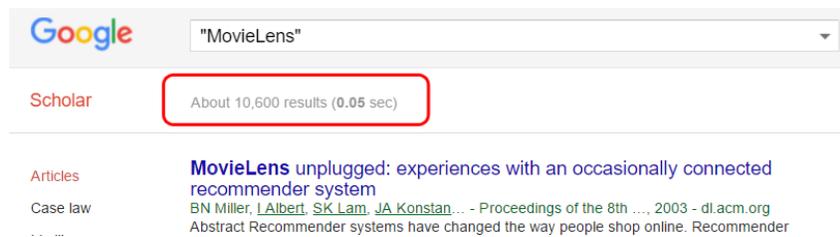

Figure 1: Number of search results on Google Scholar for "MovieLens"[3]

Sowiport is one of the largest digital libraries for the social sciences in Germany. It enables users to search in a corpus of 9.5 million research articles. Each article's detail page features a list of related articles (**Figure 2**). Sowiport, however, does not operate its own recommender system to calculate article relatedness. Instead, Sowiport uses Mr. DLib, which is a recommendation-as-a-service provider. **Figure 3** illustrates the recommendation process: (1) A partner of Mr. DLib, such as Sowiport, requests a list of related articles for a source article that is currently browsed by a user. The request is sent as a HTTP GET request to Mr. DLib's RESTful Web Service. (2) Mr. DLib calculates a list of related articles and returns the articles' metadata in XML format. (3) The partner displays the articles on its website, and (4) when a user clicks on a recommendation, the partner sends a logging notification to Mr. DLib, so that Mr. DLib knows which recommendations were clicked by the users. To enable the calculation of recommendations, Sowiport provided Mr. DLib with metadata of its 9.5 million documents (title, abstracts etc.).

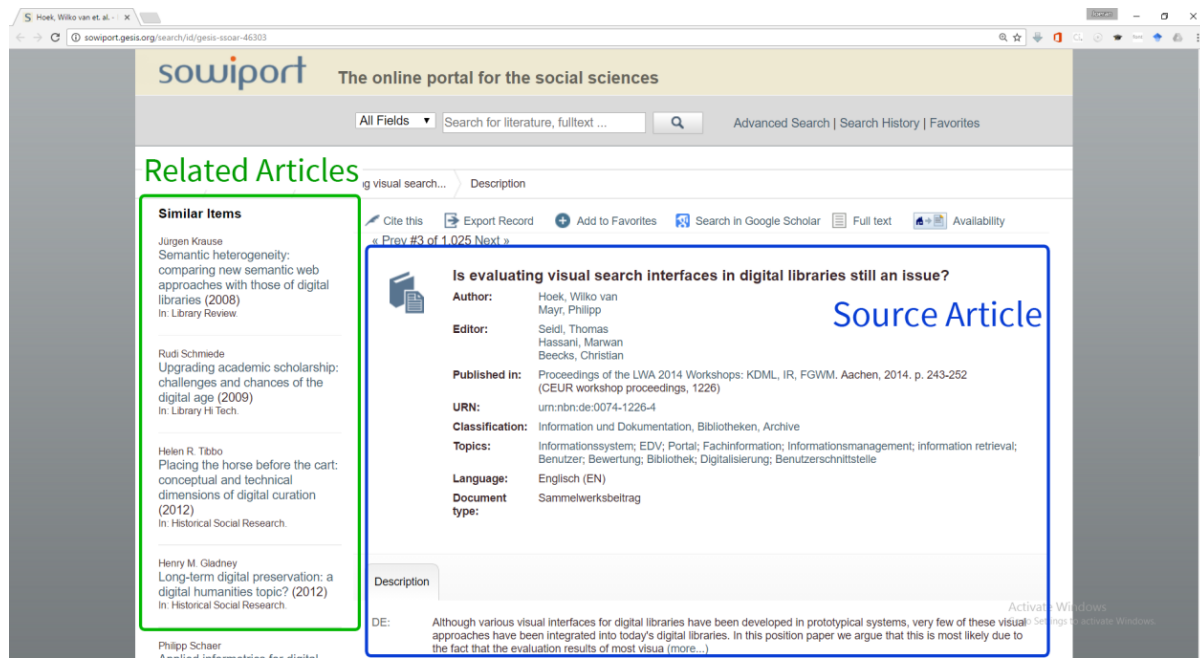

Figure 2: Screenshot of an article's detail page on Sowiport and related-article recommendations

---

Between 18th of September 2016 and 8th of February 2017, Sowiport sent 7.36 million requests for related-article recommendations to Mr. DLib. Mr. DLib returned 57.4 million recommendations for related-articles (7.8 on average per request), of which 77,468 recommendations were clicked.

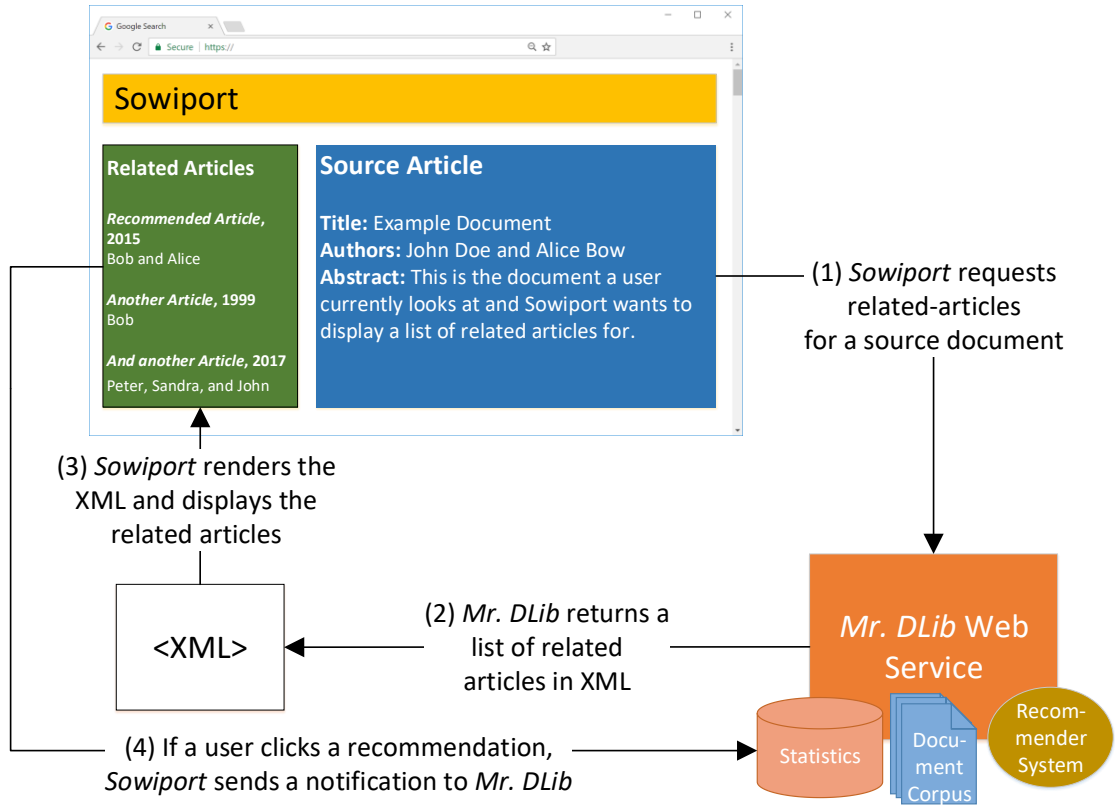

Figure 3: Illustration of the recommendation process

In the remainder, we provide background information on Mr. DLib, related work, and describe the dataset including potential use cases.

## 2 MR. DLIB

Whenever a partner requests recommendations, Mr. DLib uses an A/B engine that randomly picks a recommendation approach and varies the approach's parameters. The current distribution among the recommendation approaches is 90% content-based filtering, 4.9% stereotyping, 4.9% most-popular recommendations, and 0.2% random recommendations [Beel, Dinesh, et al. 2017].[4]

Content-based filtering (**CBF**) is implemented via Apache Solr/Lucene's[5] More-like-this function that uses BM25. Before Mr. DLib can apply CBF, partners submit their documents' metadata to Mr. DLib, and Mr. DLib indexes the data in Lucene. Metadata includes title, abstract, keywords, author names, and journal names.

---

[4] We picked this distribution, because content-based filtering is currently the best performing approach, and the A/B engine varies many parameters for the content-based filtering recommendation approach, so we want a higher proportion for the content-based filtering recommendations than for the other approaches.

[5] We use Solr version 5.5.3.

Sowiport provided Mr. DLib with metadata of 9.5 million documents, of which 56% are English, 22% German, 15% unknown, and 7% other languages. 40% of the documents have an abstract, 60% do not have an abstract. Most documents are journal articles (61%) and books (24%) (**Table 1**).

**Table 1: Document types in the Sowiport corpus**

| Journal Article | Book | Conference Article | Monograph | Website | Other | Unknown |
|---|---|---|---|---|---|---|
| 61.56% | 24.13% | 3.79% | 3.68% | 0.07% | 5.06% | 1.71% |

To enhance Lucene's standard CBF, we experiment with key-phrases that we extract from English documents' metadata and index in Lucene. Key-phrases are stemmed nouns, on which part-of-speech tagging and some statistical analyses are applied [Ferrara et al. 2011]. Overall, we calculated 13 million unique key-phrases for 5.3 million English documents. For most documents (61.5%) Mr. DLib could extract between 2 and 20 key-phrases (**Figure 4**). The maximum key-phrase count for one document is 118.

When the A/B engine chooses content-based filtering, it randomly selects whether to use "normal terms" or key-phrases for calculating document similarity. When key-phrases are chosen, the engine randomly selects if key-phrases from the 'abstract' or 'title and abstract' are used. Subsequently, the system randomly selects if unigrams, bigrams, or trigrams are used as key-phrases. Finally, the system randomly selects if one, two, ... or twenty key-phrases are used to calculate document relatedness.

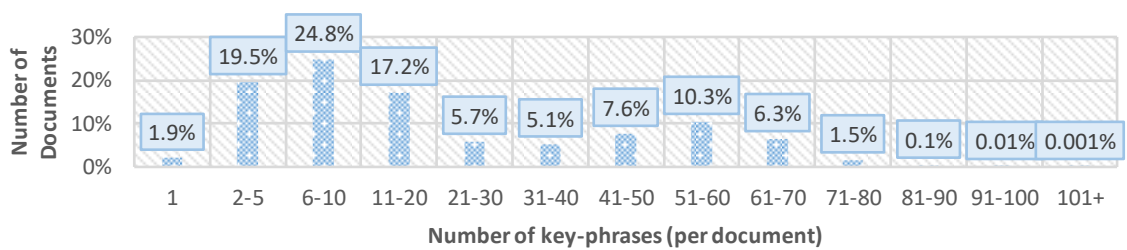

**Figure 4: Number of key-phrases**

The stereotype approach is adopted from the reference manager Docear [Beel et al. 2015]. To create stereotype recommendations, we assume that most Sowiport users – who are primarily students and researchers – are interested in the topics "academic writing", "research methods", and "peer review & research evaluation". We used Sowiport's search function to find 16 documents that we considered relevant for the three topics. These documents are flagged in Mr. DLib's database, and recommended if the stereotype approach is chosen by the A/B engine.

The most-popular approach recommends the most-popular documents from Sowiport. "Popularity" is measured by "Views", i.e. the most viewed articles on Sowiport's website, and by "Exports", i.e. the most exported documents on Sowiport's website.

If the A/B engine selects "random recommendations", the system randomly picks documents from the corpus. The random recommendations only serve as a baseline.

Once a list of recommendation candidates is created, the A/B engine chooses if it will re-rank the candidates based on bibliometrics/altmetrics that we calculate with readership statistics from Mendeley [Siebert et al. 2017]. This means, from the top $x$ candidates, those documents with the highest bibliometric relevance scores are eventually recommended. The readership data is not included in RARD, but probably will be in a subsequent release.

After the algorithm is assembled, and a list of recommendation candidates has been identified and re-ranked, the A/B engine randomly selects how many recommendations to return to the partner (1...15), and if it will shuffle the recommendations in the set. We speak of a "recommendation set" to describe the number of recommendations that Mr. DLib returns for a request by Sowiport.

The A/B engine records all details about the algorithms, as well as additional information such as when a request was received, when the response was returned, and if a recommendation was clicked. All this information is contained in the dataset. For a more detailed description of Mr. DLib's architecture and algorithms, please refer to Beel, Aizawa, et al. [2017], Beel and Dinesh [2017a; 2017b], Beel, Dinesh, et al. [2017], Beierle et al. [2017], Feyer et al. [2017] and Siebert et al. [2017].

## 3   RELATED WORK

As mentioned, there are recommendation datasets in many disciplines such as movies, books, and music. Some datasets contain only a few thousand ratings of some hundreds of users; other datasets contain millions of ratings (cf. **Table 2**). The information-retrieval community has made significant efforts to create datasets and test corpora for enabling the evaluation of information-retrieval systems, especially in the field of search. Some examples include TREC [Harman 1992; Voorhees and Ellis 2016], NTCIR [Aizawa et al. 2013; Zanibbi et al. 2016], and CLEF [Hopfgartner et al. 2016; Koolen et al. 2016] under whose umbrella several datasets were released. For academic search, the 'TREC OpenSearch – Academic Search Edition' provides datasets and a living lab that allows for the evaluation of academic search algorithms [Balog et al. 2016].

**Table 2: Key characteristics of some popular datasets**

| Dataset | Users | Items | Ratings | Density | Rating Scale |
|---|---|---|---|---|---|
| Movielens 1M | 6,040 | 3,883 | 1,000,209 | 4.26% | [1-5] |
| Movielens 10M | 69,878 | 10,681 | 10,000,054 | 1.33% | [0.5-5] |
| Movielens 20M | 138,493 | 27,278 | 20,000,263 | 0.52% | [0.5-5] |
| Jester | 124,113 | 150 | 5,865,235 | 31.50% | [-10, 10] |
| Book-Crossing | 92,107 | 271,379 | 1,031,175 | 0.0041% | [1, 10], and implicit |
| Last.fm | 1,892 | 17,632 | 92,834 | 0.28% | Play Counts |
| Wikipedia | 5,583,724 | 4,936,761 | 417,996,366 | 0.0015% | Interactions |
| OpenStreetMap (Azerbaijan) | 231 | 108,330 | 205,774 | 0.82% | Interactions |
| Git (Django) | 790 | 1,757 | 13,165 | 0.95% | Interactions |

Source: http://www.kdnuggets.com/2016/02/nine-datasets-investigating-recommender-systems.html

In the field of scientific recommender systems, there are rather few attempts to publish datasets. Lykke et al. published the iSearch collection, a dataset with around 450 thousand monographs, and journal articles, including a classification and relevance assessment for 65 topics [Lykke et al. 2010]. Sugiyama and Kan released two small datasets[6], which they created for their academic recommender system [Sugiyama and Kan 2010]. The datasets include some research papers, and the interests of 50 researchers. Gipp et al. [2015] created CITREC, a framework and dataset for evaluating recommendation algorithms for scientific literature. They use 170,000 open access articles in the life sciences and provide two gold standards – one derived automatically from subject descriptors assigned to the articles and one created manually in an earlier TREC task. They also provide source code and the results of 35 citation-based and text-based recommendation algorithms.

Several reference managers published datasets that can be used for recommender-system research. CiteULike[7] and Bibsonomy[8] published datasets containing the social tags that their users added to research articles. Although not always intended for it, the datasets are often used to evaluate research-paper recommender systems [Caragea et al. 2013; Dong et al. 2009; He et al. 2010; Huang et al. 2012; Kataria et al. 2010; Pennock et al. 2000; Rokach et al. 2013; Torres et al. 2004; Zarrinkalam and Kahani 2013]. Similarly, Jack et al. compiled a dataset based on the reference-management software Mendeley [Jack et al. 2012]. The dataset includes 50,000 randomly selected personal libraries from 1.5 million users. These 50,000 libraries contain 4.4 million articles with 3.6 million of them being unique. These datasets can be used for evaluating, for example, (item-based) collaborative filtering algorithms. The developers of the reference-management software Docear published a dataset from their recommender system [Beel et al. 2014]. The dataset contains metadata of 9.4 million academic articles, the citation network, information about 8,000 users and their 52 thousand personal libraries, and details on 300,000 recommendations that the recommender system delivered. The existing datasets are frequently used and discussed, for instance by Abbasi and Frommholz [2015], Hagen et al. [2016], White [2017], Michelbacher et al. [2011] and Stiller et al. [2014].

---

[6] http://www.comp.nus.edu.sg/~sugiyama/SchPaperRecData.html

[7] http://www.citeulike.org/faq/data.adp

[8] https://www.kde.cs.uni-kassel.de/bibsonomy/dumps/

## 4  THE DATASET

The RARD dataset is somewhat similar to the dataset of Docear [Beel et al. 2014]. To the best of our knowledge, the Docear dataset was the first dataset (and until the release of RARD also the only dataset) that contains information about how recommendations being created with variations of different recommendation approaches, and that were delivered to real users. The RARD dataset contains similar information. However, the Docear dataset focused on the user modelling process based on mind-maps [Beel et al. 2015], which was very specific to the Docear software. The RARD dataset focuses on related-article recommendation in general, and is hence potentially relevant for a broader audience. We do not claim that RARD is better (or worse) than the Docear and other existing datasets. Each of the existing datasets has its unique features, and may be useful in some scenario. We consider RARD to be an additional dataset in the domain of scientific recommender systems that may complement the existing datasets.

RARD is available on Harvard's Dataverse http://data.mr-dlib.org, and published under "Creative Commons Attribution 3.0 Unported (CC-BY)" license[9]. If a user feels that our license is not appropriate for his or her purposes, we are open to discuss alternatives. Due to copyright restrictions, we are not able to publish the documents' metadata. However, we publish a list of the original document IDs, which allows for the request of metadata for many of the documents from Sowiport's OAI-PMH interface[10].

The RARD dataset consists of tab-separated plain-text CSV files that relate to 1) the recommendation logs, 2) an implicit item-item rating matrix, and 3) a list of external IDs to lookup additional information. Each of the three sub-datasets is explained in the following.

### 4.1  Recommendation Logs and Statistics

The file `recommendation_log.csv` (7 GB) contains information about 7,360,026 related-article requests from Sowiport, and the 7.36 million corresponding recommendation sets that Mr. DLib returned. The 7.36 million recommendation sets contained 57,435,086 recommendations in total (7.8 recommendations on average). Each of the 57.4 million recommendations is represented by one row in the dataset. For each row, the dataset provides the following details/columns:

- ▪ `recommendation_id`: A unique ID for each delivered recommendation.
- ▪ `recommendation_set_id`: A unique ID of the recommendation set in which the recommendation was delivered. If one set contained, for instance, seven recommendations, then the dataset will contain seven rows in which the `recommendation_set_id` is identical.
- ▪ `set_size`: The number of recommendations in the recommendation set.

---

- `rank_in_set`: The final rank of the recommendation in the recommendation set.
- `source_document_id`: ID of the document for which related-articles where requested, i.e. the document that a user browsed on Sowiport's website.
- `recommended_document_id`: The ID of the document that was returned as recommendation.
- `request_received`: Time when Mr. DLib received the request for related-article recommendations.
- `response_delivered`: Time when Mr. DLib returned the recommendations.
- `clicked`: Time when the recommendation was clicked; `NULL` if not clicked.
- `algorithm-class`: The main recommendation category, i.e. content-based filtering (`cbf`), most-popular (`mstp`), stereotype (`strtp`) or random (`rndm`).
- `algorithm_id`: ID of the specific algorithm to calculate recommendations.
- `cbf-document_field`: The document field used to create content-based recommendations (`title` | `title_and_abstract`). `NULL` if an algorithm-class other than `cbf` was used.
- `cbf-feature_type`: The type of feature that was used [`terms` | `unigram-key-phrases` | `bigram-key-phrases` | `trigram-key-phrases`]. `NULL` if an algorithm-class other than `cbf` was used.
- `cbf-key-phrase_count`: The number of key-phrases being used [`1…20`]. `NULL` if key-phrases were not used.
- `text_relevance_score`: Lucene's text relevance score. `NULL` if an algorithm-class other than `cbf` was used.
- `popularity_metric`: The metric in which popularity is measured [`top_views` | `top_exports`]. `NULL` if an `algorithm-class` other than `mstp` was applied.
- `stereotype_category`: The stereotype category being applied [`academic-writing` | `research-methods` | `peer-review`]. `NULL` if an algorithm-class other than `strtp` was applied.

Overall, recommendations were requested at least once for 2,176,984 documents (23%) of the 9.5 million documents in the corpus. 77,468 of the 57 million recommendations were clicked, which results in a click-through rate (CTR) of 0.135%. A CTR of 0.135% may seem low compared to click-through rates of 5% and higher in, for instance, Docear [Beel 2015; Beel et al. 2014]. However, we expected low CTRs for three reasons. First, Mr. DLib also delivers recommendations when web crawlers index the website of Sowiport. This means, an unknown fraction of the 57 million recommendations was not shown to humans but to bots. In contrast, clicks are logged with JavaScript, which is usually not executed by web crawlers. Consequently, the number of recommendations contains some noise, while the click counts should be mostly reliable. Second, recommendations on Sowiport typically open in a new browser tab in the foreground. This means, users who want to click a second recommendation need to go back to the original tab. This is a rather high burden and we assume that many users do not do this. Third, Sowiport shows related-article recommendations on every article's detail page. Consequently, users looking at several articles, receive many recommendations, which might lead to a decreased attention level for recommendations after a while.

The recommendation logs enable researchers to conduct research on data from a production recommender system, without implementing a recommender system themselves, and without doing any user studies. Researchers could re-do the analyses we already did [Beel, Aizawa, et al. 2017; Beel and Dinesh 2017a; Beel and Dinesh 2017b; Beel, Dinesh, et al. 2017; Beierle et al. 2017] but with more advanced statistical methods. For instance [11], researchers can calculate click-through rates for content-based filtering (0.145%), stereotype recommendations (0.124%), most-popular recommendations (0.11%), and random recommendations (0.12%), and conclude that content-based filtering is the best performing recommendation approach in Mr. DLib. This analysis would be performed by counting the delivered recommendations, i.e. the rows in the dataset where `algorithm_class` equals `cbf`, `strtp`, `mstp`, and `rndm` respectively. Clicked recommendations are identified by counting the rows where `recommendation_class` equals `cbf`, `strtp`, ... `AND clicked != NULL`. Researchers could also explore the time Mr. DLib needed to generate recommendations. They would only have to subtract `time_request_received` from `time_response_delivered` for each row. This would lead to a chart as shown in **Figure 5**, and show that 34% of Mr. DLib's recommendations required between one and two seconds to calculate.

---

[11] All examples provided in this section have already been published. The purpose of describing them here again is to demonstrate what the dataset can be used for.

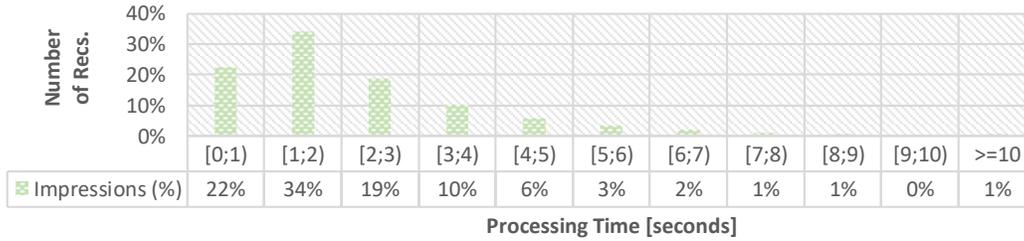

**Figure 5: Time to calculate recommendations**

Researchers could also conduct completely new analyses. Some analyses we have not yet conducted include research on position bias (analyze if and to what extent recommendations shown first in the recommendation list are clicked more often than documents at the lower ranks); analyze if normal terms or key-phrases are more effective, and which parameters of key-phrases should be chosen (e.g. uni-grams bi-grams, or tri-grams; one, two, ... or twenty key-phrases); analyze if content-based filtering based on the title, or title and abstract is more effective; and analyze how well Lucene text relevance scores and click-through rates correlate.

## 4.2    Implicit Rating Matrix

The file `rating_matrix.csv` (429 MB) contains 2,176,984 million rows. Each row represents a source document, i.e. a document for which Sowiport requested recommendations, and the document IDs of the returned related-article recommendations. The file represents an implicit item-item rating matrix, which was calculated based on the recommendation log. The matrix was calculated based on the assumption that a click [non-click] on a recommendation can be considered as implicit positive [negative] rating of how relevant the recommendation was, given a specific source document.

**Table 3: Implicit Rating Matrix (Illustration)** [12]

| source_document_id | Triplet 1 | | | Triplet 2 | | | ... | Triplet i | | |
| --- | --- | --- | --- | --- | --- | --- | --- | --- | --- | --- |
| | recommended_document_id | Click Count | Non-Click Count | recommended_document_id | Click Count | Non-Click Count | ... | recommended_document_id | Click Count | Non-Click Count |
| 2 | 1 | 2 | 1 | 8 | 1 | 0 | ... | 143 | 1 | 13 |
| 5 | 534555 | 1 | 1 | NULL | NULL | NULL | ... | NULL | NULL | NULL |
| 6 | 42334 | 1 | 0 | 5343 | 1 | 2 | ... | NULL | NULL | NULL |
| 7 | 3 | 0 | 2 | 4 | 5 | 0 | ... | 7565 | 5 | 0 |
| 12 | 34 | 2 | 0 | 423 | 0 | 8 | ... | 12 | 12 | 43 |
| 14 | 533 | 0 | 4 | NULL | NULL | NULL | ... | NULL | NULL | NULL |
| 15 | 423 | 1 | 8 | 5521 | 1 | 1 | ... | NULL | NULL | NULL |
| ... | ... | ... | ... | ... | ... | ... | ... | ... | ... | ... |
| n | 6652 | 2 | 4 | NULL | NULL | NULL | ... | NULL | NULL | NULL |

The concept is illustrated in **Table 3**. The first column contains the source document ID (`source_document_id`). The following columns contain `i` triples, one for each document that Mr. DLib recommended as related-article for the source document. Each triple contains a) the ID of the recommended document (`recommended_document_id`), b) a counter of how often the document was recommended as related to the source document and clicked (`Click Count`), and c) a counter how often the document was recommended as related to the source document and not clicked (`Non-Click Count`). For instance, based on **Table 3**, Sowiport has requested recommendations for document ID=15 several times (the precise number cannot be inferred from this dataset but from the recommendation log). Mr. DLib has returned document ID=423 nine times as recommendations, and from these nine times, the document has (not) been clicked 1 (8) times. Mr. DLib also returned document 5521 as recommendation twice, whereas this recommendation was once clicked, and once not clicked.

---

[12] The table contains NULL values for illustrational purposes. The actual dataset contains only the NOT NULL values.

### 4.3 Filtered Implicit Rating Matrix

When interpreting non-clicks as negative ratings, it must be kept in mind that an unknown amount of recommendations was delivered to web crawlers, i.e. the non-clicks contain noise. Therefore, we created a second implicit item-item rating matrix, and excluded recommendation sets in which not at least one recommendation was clicked. The rationale is that when no recommendation in a set was clicked, we do not know if the recommendations were just not relevant, or if recommendations were delivered to a web crawler. In contrast, when at least one recommendation was clicked in a set, we assume that a real user has looked at the recommendations and decided to click at least one of them. Consequently, also the non-clicks in these sets are more meaningful. The resulting matrix is provided in the file `rating_matrix_fitered.csv` (7 MB). It contains 63,923 rows, i.e. unique source documents for which recommendations were delivered and at least one recommendation in the set was clicked.

### 4.4 External IDs (Sowiport, Mendeley, arxiv, ...)

The file `external_IDs.csv` (280 MB) contains Mr. DLib's document IDs, the original document IDs from Sowiport (100%), Mendeley IDs (18%), ISSN (16%), DOI (14%), Scopus IDs (13%), ISBN (11%), PubMed IDs (7%) and arxiv IDs (0.4%). The file `external_IDs.csv` allows for the lookup of further information (e.g. metadata) on the documents from Sowiport's API[13], Scopus, PubMed, Mendeley, etc. All IDs, except Sowiport IDs, were retrieved from Mendeley's API[14] by sending document titles to the API and parsing the response. Some documents might be assigned with incorrect IDs (except Sowiport IDs) because Mendeley only had the title for document identification. Hence, it might be that Mendeley returned IDs for a document that was not the same document as in our database, but only had the same, or very similar, title. Mendeley's data is also available under the CC-BY license.

## 5 LIMITATIONS AND OUTLOOK

We introduced RARD, a novel dataset in the domain of scientific recommender systems. RARD contains information about 57,435,086 million recommendations of which 77,468 were clicked. The dataset can be used for various analyses of Mr. DLib's recommendation algorithms, to evaluate collaborative filtering recommendation approaches, and train machine learning recommendation algorithms. Clicks are only a weak measure of relevance, and the number of clicks in the dataset is rather low (77 thousand), while the number of items is high (9.5 million). Hence, we do not claim that our dataset is as good as, for instance, the MovieLens dataset with millions of explicit user ratings. However, in the field of scientific recommender systems, our dataset is quite unique. To the best of our knowledge, there is no dataset with more (implicit) ratings available, and with as many variations of recommendation algorithms.

We plan to release annual updates of the dataset, and to address the following issues.

1. Data quality
   The biggest limitation of our dataset is that we do not detect web crawlers. Hence, the value of the non-clicks is reduced as are statements about how many recommendations Mr. DLib delivers. To overcome that limitation, we currently implement a logging mechanism based on JavaScript. This allows for identifying non-clicks from real users, while excluding web crawlers.

---

[13] http://sowiport.gesis.org/OAI/Home

[14] http://dev.mendeley.com/

2. Data quantity and variety

We aim at delivering more data and variety in the next dataset. We plan to implement new algorithms e.g. to enable cross-lingual recommendations [Zhou et al. 2012], enrich meta data [Stiller et al. 2014], vary more parameters, include readership statistics from Mendeley, record more statistics, import an additional 20 million open access documents [Knoth and Pontika 2016; Knoth and Zdrahal 2012; Pontika et al. 2016], and integrate Mr. DLib in two new partners' products, one of them being JabRef, one of the most popular open-source reference managers [Feyer et al. 2017]. In addition, we currently discuss to connect external recommender systems such as the CORE recommender [Knoth et al. 2017], to Mr. DLib. This means, the next release of RARD will provide more data in terms of more content, more algorithms, more partners, and more recommendations and clicks.

3. Scope

In the long run, Mr. DLib will not only recommend research-articles, but also other items relevant in Academia, such as citations [Chakraborty et al. 2015], reviewers [Wang et al. 2010], call for papers [Beierle et al. 2016], research grants, and university courses.

4. Document Metadata

We hope that in our next dataset, we can include the documents' metadata. We believe this would greatly enhance the value of the dataset and allow many additional analyses. However, this eventually depends on our partner's licenses. Currently we discuss this option with the operator of Sowiport. Nevertheless, we want to emphasize that the documents' metadata is mostly already available through Sowiport's OAI interface, and additional usage data is available [Mayr 2016]. Interested researchers can hence download the documents' metadata and usage data and link that data with our dataset.

5. Relevance feedback beyond clicks

Currently, Mr. DLib only records clicks, i.e. implicit relevance feedback. We plan to enhance Mr. DLib and allow users to directly rate the quality of recommendations on a five-star scale. In addition, we plan to record more meaningful relevance feedback such as if a recommended document was exported, printed, or added to the user's favorites. One of the two new partners (JabRef) will also enable us to monitor if a recommended document was added to a user's collection.

6. User sessions

We plan to enable the logging of user sessions. In that case, we would not only know that a set of documents was recommended (and clicked) for a particular source document, but which source documents were viewed by a user within a session, and which number of recommendation sets and recommendations were displayed to a user during a session. Such data will allow us to implement more sophisticated recommendation approaches, more detailed analyses of the users' behavior, and creating a more comprehensive rating matrix.

7. Personalized recommendations

Currently, Mr. DLib focuses on related-article recommendations. In the long-run, we aim at providing personalized recommendations, maybe even across different partner websites [Koidl et al. 2013]. This requires Mr. DLib to access users' personal document collections, which is not only technically more challenging to implement, but also challenging from a data privacy perspective.

Finally, we would like to emphasize that we are open for cooperation and to include further data sources into the dataset. Therefore, if you are interested in a cooperation, using the dataset, or if you have questions, please do not hesitate to contact us.

## 6 AUTHORS

Joeran Beel is Assistant Professor in Intelligent Systems at Trinity College Dublin, and member of the ADAPT Center. His research focuses on recommender systems: recommendations as a service, recommender-system evaluation, and user modelling in the domain of digital libraries. Joeran has published three books and over 50 peer-reviewed articles, and has been awarded various grants for research projects, patent applications, and prototype development as well as some business start-up funding. He is involved in the development of several open-source projects such as Docear, Mr. DLib, JabRef, and Freeplane, some of which he initiated. Joeran founded two successful IT start-ups and received multiple awards and prizes for each.

Zeljko Carevic is a research associate at GESIS – Leibniz Institute for the Social Sciences in Cologne Germany. Since 2015 he is a Ph.D. candidate at the University of Duisburg-Essen. His research focuses on exploratory search in digital libraries with a special interest in bibliometrics, citation analyses as well as personalization and contextualization incorporated in Sowiport.

Johann Schaible is postdoctoral researcher and team leader of team Knowledge Discovery at the GESIS Computational Social Science department. His research focuses on Semantic Web, Linked Data, and recommender systems, specifically based on association rule mining and learning to rank approaches. Johann published over ten peer-reviewed articles at leading international conferences (e.g. Extended Semantic Web Conference) as well as journals (e.g. Semantic Web Journal) and is currently involved in several grant proposals considering Open Access transformation in digital libraries.

Gábor Neusch obtained his MSc degree in Business Information Technology at the Corvinus University of Budapest in 2014. At the same year, he started his PhD studies in the same field at the Department of Information Systems of Corvinus. His research area is knowledge management, more precisely ontology based adaptive learning solutions. The focus of his work lies on ontology mapping at the time being. He works as a consultant and researcher for the Future Internet Living Lab Association.

## ACKNOWLEDGEMENTS

This publication has emanated from research conducted with the financial support of Science Foundation Ireland (SFI) under Grant Number 13/RC/2106. We are further grateful for the support by Philipp Meyer, Sophie Siebert, Siddharth Dinesh, and Andrew Collins.